\documentclass[preprint,12pt]{elsarticle}

\usepackage{amssymb}

\journal{Physics Letters B}

\begin{document}

\begin{frontmatter}



\title{Neutral-current neutrino cross section and expected supernova signals for $^{40}$Ar from a three-fold increase in the magnetic dipole strength}

\author[Duke,TUNL]{W. Tornow}
\author[LLNL,Duke]{A. P. Tonchev}
\author[Duke,TUNL]{S.W. Finch}
\author[Duke,TUNL]{Krishichayan}
\author[Huzhou]{X. B. Wang}
\author[LANL]{A. C. Hayes}
\author[LANL,Kubrick]{H. G. D. Yeomans}
\author[LANL,MIT]{D. A. Newmark}

\affiliation[Duke]{Department of Physics, Duke University, Durham, NC 27708-0308, USA}
\affiliation[TUNL]{Triangle Universities Nuclear Laboratory, Durham, NC 27708-0308, USA}
\affiliation[LLNL]{Nuclear and Chemical Sciences Division, Lawrence Livermore National Laboratory, Livermore, CA 94550, USA}
\affiliation[Huzhou]{School of Science, Huzhou University, Huzhou 313000, China}
\affiliation[LANL]{ Los Alamos National Laboratory, Los Alamos, New Mexico 87545, USA}
\affiliation[Kubrick]{Kubrick Group, London, SE1 0BE, United Kingdom}
\affiliation[MIT]{Department of Physics, Massachusetts Institute of Technology, Cambridge, Massachusetts,02139, USA}

\begin{abstract}
In view of the great interest in liquid argon neutrino detectors, the \textsuperscript{40}Ar($\gamma,\gamma'$)\textsuperscript{40}Ar\textsuperscript{*} reaction was revisited to guide a calculation of the neutral current neutrino cross section at supernova
energies. Using the nuclear resonance fluorescence technique with a monoenergetic, 99\% linearly polarized photon beam, we report a three-fold increase in magnetic dipole strength at around 10 MeV in  $^{40}$Ar. Based on shell-model calculations, and using the experimentally identified transitions, the neutral current neutrino cross sections for low-energy reactions on \textsuperscript{40}Ar are calculated.
\end{abstract}



\begin{keyword}



\end{keyword}

\end{frontmatter}



\section{Introduction}

Many current and planned neutrino facilities are based on liquid-argon (LAr) detector designs, for example DUNE \cite{DUNE}, the Deep Underground Neutrino Experiment, at the Sanford Underground Research Facility (SURF) in South Dakota \cite{SURF}. While the focus of these neutrino platforms is state-of-the-art neutrino oscillation studies, most LAr projects are also poised to detect neutrinos from core-collapse supernovae (SN).
Although the charged-current (CC) signal for SN neutrinos in LAr detectors has been studied \cite{Garcia} via the reactions
$$\nu_e + ^{40}\mathrm{Ar} \rightarrow ^{40}\mathrm{K}^* + e^- $$
and 
$$ \bar{\nu}_e + ^{40}\mathrm{Ar} \rightarrow ^{40}\mathrm{Cl}^* + e^+ \, , $$
information on a neutral-current (NC) signal in the energy region of interest is very limited.
In this Letter we concentrate on the nuclear excitation of \textsuperscript{40}Ar nuclei in LAr by the NC neutrino interaction
$$\nu + ^{40}\mathrm{Ar} \rightarrow \nu' + ^{40}\mathrm{Ar}^*,$$
and the corresponding, de-excitation $\gamma$-ray signals, which 
could shed light on the physics of core-collapse SN 
neutrino bursts. The measured M1 and E1 strength was used to estimate the CC and NC cross sections for the neutrino induced reactions on $^{40}$Ar.

Li \textit{et al.} \cite{Li} reported the first observation of a spin-flip magnetic dipole (M1) transition in \textsuperscript{40}Ar at 9.757 MeV 
with an assigned strength of $B(M1\uparrow)$=0.148(59)$\mu_N^2$. 
Shell-model calculations \cite{Li} suggested this state was one fragment of the spin-flip M1 strength in \textsuperscript{40}Ar. 
Recently, Gayer {\it et al.} \cite{Gayer}, studied dipole and quadrupole excited states of $^{40}$Ar between 4.2 and 7.7 MeV. In LAr detectors the de-excitation $\gamma$ rays are detected via electron-positron pair production, favoring $\gamma$ rays of energies above the energy range studied in Ref.~\cite{Gayer}. 
The large uncertainty associated with the $B(M1\uparrow)$ value in \cite{Li} and the possibility of locating additional M1 strength around 10 MeV motivated us to revisit the \textsuperscript{40}Ar($\gamma,\gamma'$)\textsuperscript{40}Ar reaction and calculate the NC cross section, which is closely related to $B(M1\uparrow)$ at very low energies \cite{Langanke}.

\section{Experimental measurement of magnetic dipole strength}

The Nuclear Resonance Fluorescence (NRF) technique \cite{NRF2,NRF1, Rus13, Ton17} was used to measure the magnetic dipole strength $B(M1\uparrow$) at the High-Intensity Gamma-Ray Source (HI$\gamma$S) \cite{HIGS} of the Triangle Universities Nuclear Laboratory (TUNL) \cite{TUNL}. 
Linearly polarized (99\%) and quasi-monoenergetic 9.88 MeV photons were produced via Compton back-scattering of 543 MeV electrons from 548 nm free-electron laser photons.
After traveling 53 m in a high-vacuum tube the photon beam passed through a 1.27 cm diameter and 15 cm long lead collimator located approximately 4 m upstream of the $^{40}$Ar target. 
The energy was centered at 9.88 MeV, with approximately 300 keV full width at half maximum. The average photon flux during the approximate 36.5-hour \textsuperscript{40}Ar irradiation was 2$\times$10$^{7} \gamma$/s. 
The \textsuperscript{40}Ar target consisted of an 84 mm inner diameter vessel made of 1 mm thick aluminum covered by a 4 mm thick layer of carbon fiber to accommodate the \textsuperscript{40}Ar pressure of 307.1 atm. 

\begin{figure}[t]
\centering
\includegraphics[width=13 cm, angle=0]{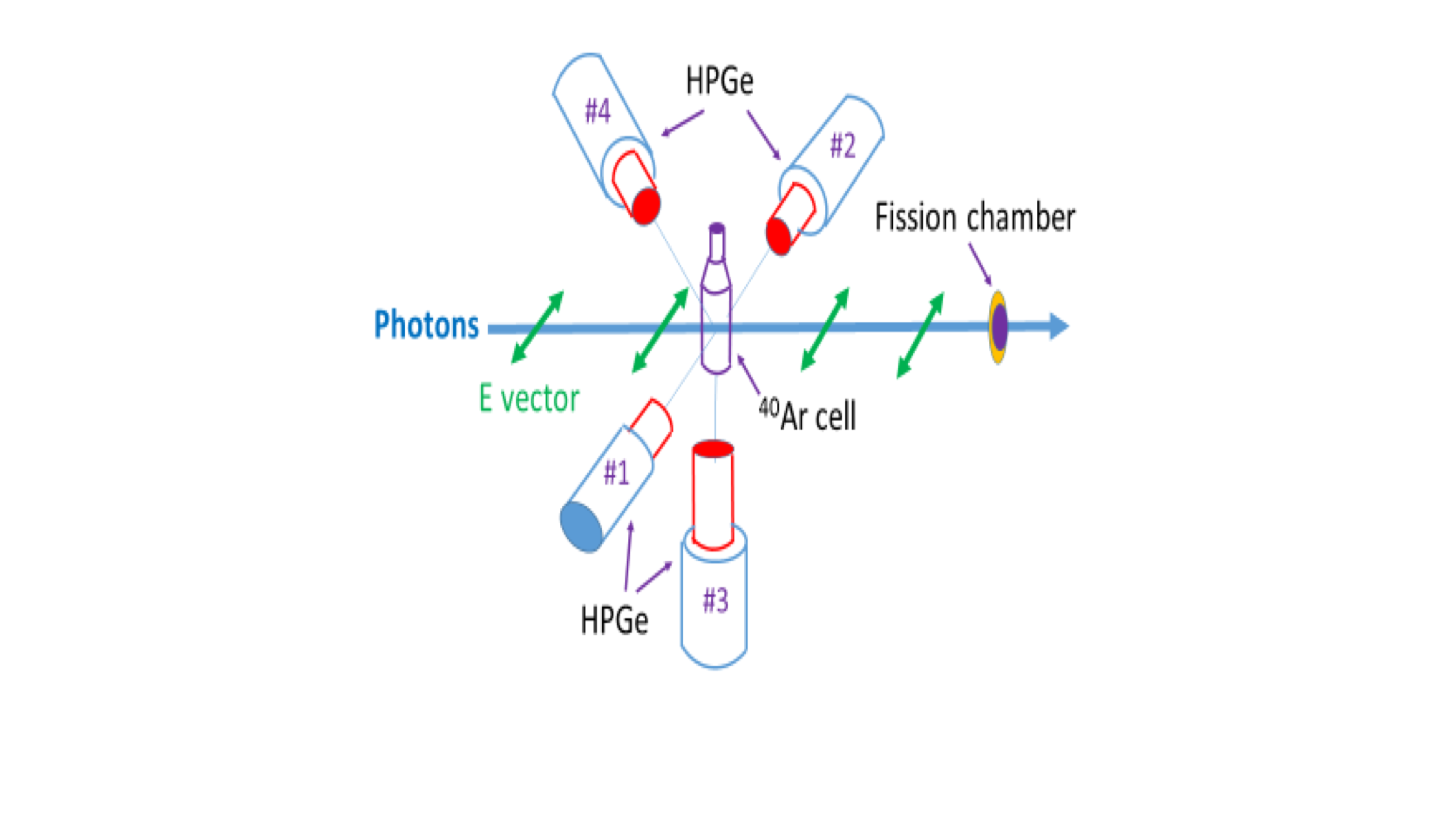}
\caption{Schematic of experimental setup. 
The linearly polarized (in the horizontal plane) photon beam enters from the left side and passes through an $^{40}$Ar filled container, which is viewed by four HPGe detectors, the angles of which are given in the text.
The photon flux is monitored by the $^{238}$U fission chamber shown on the right side.                         
}
\label{two}
\end{figure}

A schematic of the experimental setup is shown in Fig. \ref{two}.
Four High-Purity Germanium (HPGe) detectors were used to record 
the scattered $\gamma$ rays from $^{40}$Ar. 
Two 60\% efficient HPGe detectors (relative to a $3^{''} \times 3^{''}$ NaI detector) were placed in the horizontal plane at polar angle $\theta=90^o$ 
on either side of the incident photon beam at azimuthal angle $\phi=0^o$ and $180^o$ to record de-excitation 
$\gamma$ rays of M1 character,
while a 100\% efficient HPGe detector was positioned in the vertical plane at 
$\theta=90^o$ below the photon beam at $\phi= 270^o$ to record de-excitation $\gamma$ rays 
of electric dipole (E1) nature. 
A fourth, 60\% efficient HPGe detector was placed at $\theta$=127$^o$ and $\phi=135^o$ 
to check on unlikely de-excitation $\gamma$ rays of electric quadrupole (E2) nature.
The distance between the center of the argon container and the front face of the HPGe detectors \#1, \#2, \#3, and \#4 (see Fig.~\ref{two}) is 13.7 cm, 12.8 cm, 13.8 cm, and 11.4 cm, respectively. The detector diameters ranged from 6.46 to 6.97 cm. 
A thin-walled, $^{238}$U-based fission-ionization chamber was positioned downstream of the $^{40}$Ar cell to record the incident photon flux \cite{fission}.

\begin{figure}[t]
\centering
\includegraphics[width=10 cm, angle=0]{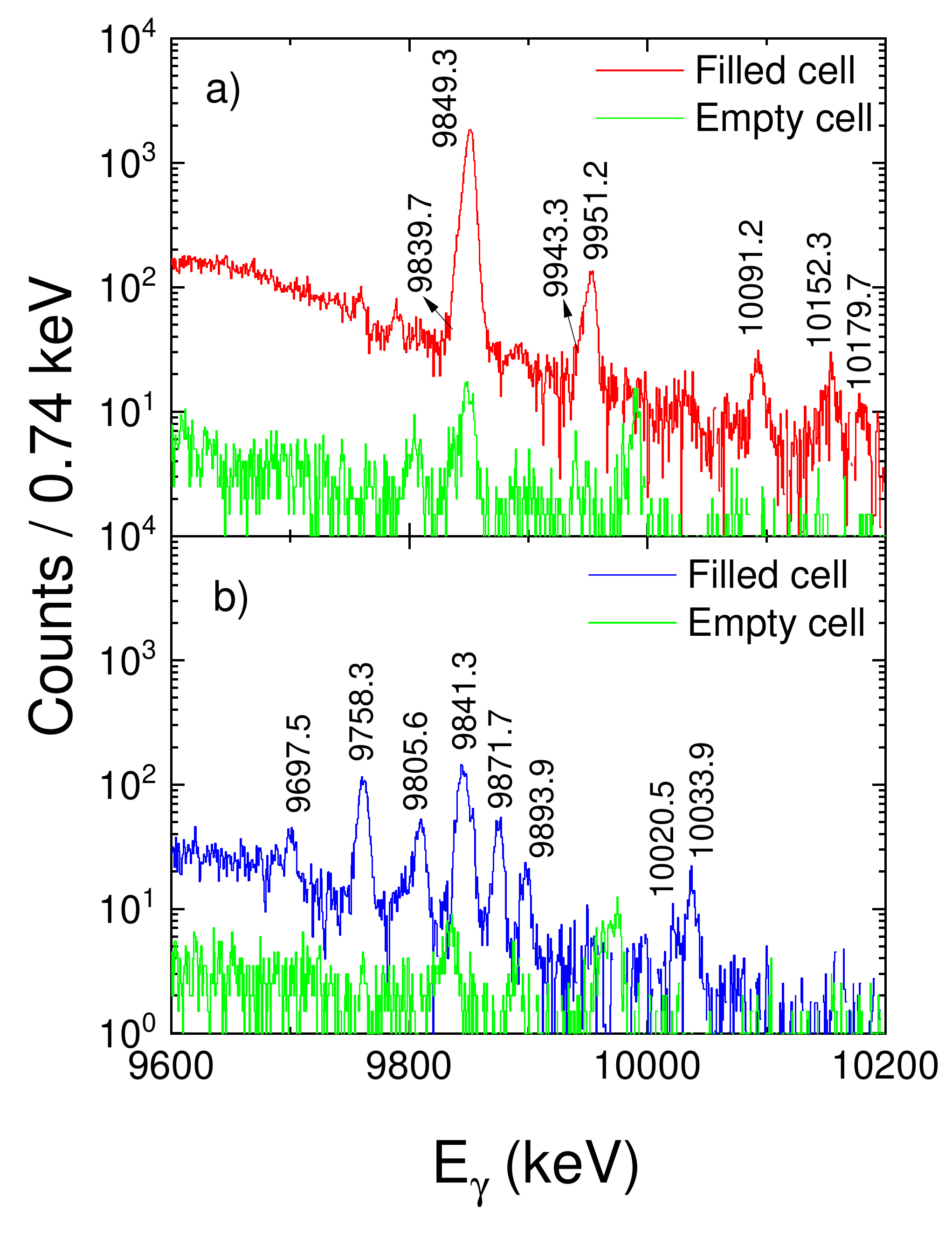}
\caption{ (a) $\gamma$-ray energy spectrum obtained with vertical HPGe detector and filled $^{40}$Ar cell (red). 
This detector records E1 de-excitation $\gamma$ rays. The peaks are identified by the excitation energies of $^{40}$Ar, as reported in Table 1.
Also shown is the normalized spectrum measured with the empty cell (green). 
(b) Same as in (a), but for the horizontal HPGe detector $\#2$, which records M1 de-excitation $\gamma$-rays. Note the logarithmic vertical scale.}
\label{three}
\end{figure}

\begin{figure}[t]
\centering
\includegraphics[width=10 cm, angle=0]{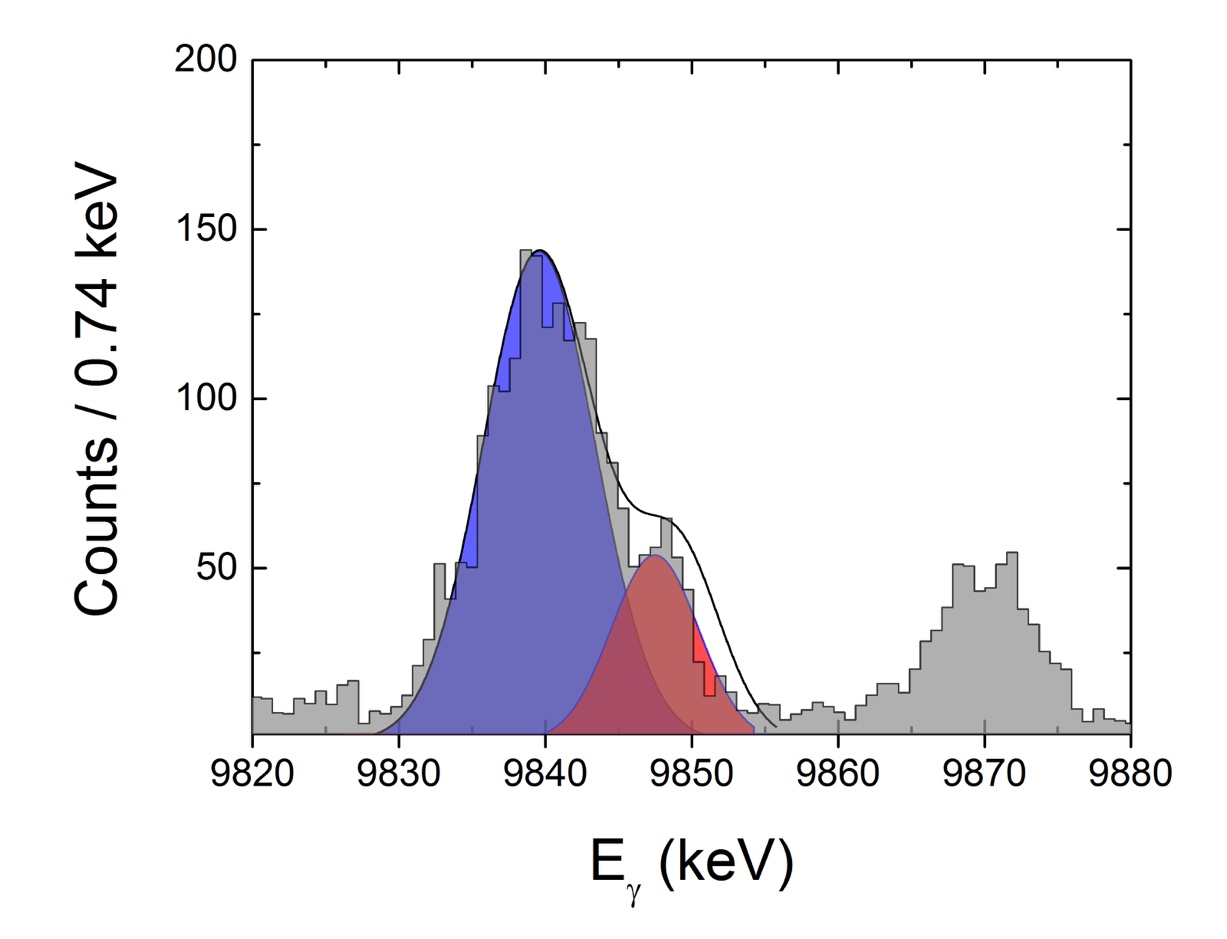}
\caption{Approximate decomposition of the $\gamma$-ray line shape observed in the 9830 keV to 9855 keV energy region with HPGe detector \#2. The Gaussian centered at 9849 keV is a contamination resulting from the strong E1 transition in $^{40}$Ar, while the Gaussian centered at 9840 keV is the M1 transition of interest.  }
\label{peak_decomp}
\end{figure}

A typical HPGe detector spectrum obtained with the vertical detector in the 9.4 to 10.2 MeV $\gamma$-ray energy range is shown in Fig. \ref{three} (a). 
This spectrum is dominated by the strong E1 transition from the decay of the 9849 keV state in $^{40}$Ar. 
Fig. \ref{three} (a) also shows a normalized $\gamma$-ray spectrum measured with the filled $^{40}$Ar cell replaced by an empty one, 
clearly indicating that there are no interfering $\gamma$-ray transitions in the region of interest resulting 
from the aluminum and/or carbon fiber of the containment vessel, the lead collimators, and the environment.  
Fig. \ref{three} (b) presents the $\gamma$-ray spectrum recorded with one of the two horizontal HPGe detectors.
This spectrum features M1 transitions in $^{40}$Ar and is richer than the associated E1 spectrum presented in Fig. \ref{three} (a).
The known M1 transition at 9757 keV is clearly seen. However, there is also a comparably strong $\gamma$-ray transition at 9840 keV, 
and weaker $\gamma$-ray lines originating from levels in $^{40}$Ar at 9697.5, 9805.6, 9871.7, 9893.9, 10020.5 and 10033.9 keV. 
A comparison of the 9840 keV $\gamma$-ray line with others, including that at 9757 keV, 
shows that the line in question is broader than those in its vicinity. 
According to Figs. \ref{three} (a) and \ref{three} (b) the energy of the very strong E1 transition of 9849 keV almost coincides with that of the M1 transition at 9.840 MeV.
Inspecting Fig. 2 of Li $\textit{et al.}$ \cite{Li} we note that the new M1 transition at 9840 keV is also seen in this work, but without mention in the text.
Monte-Carlo calculations performed for the present geometrical arrangement of the $^{40}$Ar target 
and the HPGe detectors predict that our measured M1 spectrum includes a contamination of approximately 3\% of the 
E1 yield, resulting in a 23\% reduction of the M1 strength at 9849 keV. 
This conclusion is in agreement with a two-Gaussian fit attempt to the line shape recorded at 9849 keV, as can be seen from Fig.~\ref{peak_decomp}. 
Focusing on the energy calibration of our HPGe detectors, we note that natural background lines, and $\gamma$-ray test 
sources, including $^{56}$Co, were used to cover the energy range up to 3.45 MeV. 
Beyond this energy, known $\gamma$-ray transitions in $^{27}$Al excited by the incident photon beam were used to extend the 
energy calibration up to 10 MeV. 
This procedure provided results consistent with previous energy assignments for states in $^{40}$Ar. 

The normalized $^{40}$Ar-out spectra were subtracted from the $^{40}$Ar-in spectra, 
with the normalization factor deduced from the yields recorded with the $^{238}$U based fission 
chamber referred to earlier.

The dipole strengths are defined as \cite{ring} 
$$
B(E1 \uparrow)=2.866 \times 10^{-3}\Gamma_0/\omega^3 \;\;\;{\rm e^2fm^2} 
$$                                                                  
and
$$
B(M1 \uparrow)= 0.2592\;\; \Gamma_0/\omega^3 \mu_n^2   \;\;\; {\rm MeV^3/meV} .                                                                                 
$$
Here, $\Gamma_o$ is the elastic scattering width and $\omega$ is the excitation energy of the state.
Inspecting our spectra, we find no evidence for inelastic transitions from the strong E1 and M1 states of interest to the 1\textsuperscript{st} excited 2$^+$ 
or other higher excited states in $^{40}$Ar. We conclude that possible inelastic processes are very small, justifying the relation $\Gamma_0 = \Gamma_f = \Gamma$ with an uncertainty of 5\%. 
$\Gamma$ and $\Gamma_f$ refer to the total and partial width for de-excitation to the excited final state $f$, respectively. Therefore, the NRF cross section is given by \cite{Hamilton} 
 
\begin{equation}
I_s=g\left(\pi\hbar/\omega\right)^2\Gamma_0
\label{cross2}
\end{equation}

with spin factor $g=(2J_x+1)/(2J_{\mathrm{g.s.}}+1)=3$ for the even-even target nucleus $^{40}$Ar with $J_{\mathrm{g.s.}}^\pi=0^+$.

Alternatively, the cross section is defined as \cite{Hamilton}
\begin{equation}
I_s=A/\left(\Phi_\gamma\epsilon_\gamma n_{Ar}W(\theta,\phi)t\right).
\label{cross3}
\end{equation}
Here, $A$ refers to the photo-peak yield, $\Phi_\gamma$ is the photon flux per second per eV, $\epsilon_\gamma$ 
is the HPGe detector photo-peak efficiency, 
$n_{Ar}$ the number of $^{40}$Ar nuclei per cm$^3$, 
$W(\theta,\phi)$ is the angular distribution function (1.5 for point geometry in case of M1) and finally, 
$t$ represents the irradiation time in seconds.

The efficiency of the HPGe detectors was determined by using calibrated $\gamma$-ray check sources with energies up to 3.45 MeV. 
The Monte-Carlo based MCNP6 \cite{MCNP} code was used to simulate and match the measured efficiency of our HPGe detectors. 
The MCNP6 calculations were then used to provide the HPGe detector efficiencies at the higher energies of interest.

The $^{40}$Ar density was obtained by weighing the empty and filled cell and by determining its inner dimensions. 
Finally, from Monte-Carlo simulations of the experimental setup the value $W=1.45$ was obtained.   
Equating Eqns. \ref{cross2} and \ref{cross3} provides the elastic scattering width $\Gamma_0$ needed to determine $B(E1\uparrow)$ and $B(M1\uparrow)$ values. 

\begin{table}[b]
\caption{Experimental results for $J^\pi=1^+$ and $J^\pi=1^-$ states from 9700 to 10200 keV of excitation in $^{40}$Ar.}
\centering
\begin{tabular}{cccc}
$\omega$ (keV)&$J^{\pi}$&$\Gamma_0$ (meV)&$B(M1\uparrow$) $(10^{-3} \mu_N^2)$\\\hline
9697.5 $\pm$ 1.4&	1$^+$&	233 $\pm$ 27&	66 $\pm$ 8\\
9758.3 $\pm$ 1.1&	1$^+$&	692 $\pm$ 60&	193 $\pm$ 17\\
9805.6 $\pm$ 1.3&	1$^+$&	272 $\pm$ 26&	75 $\pm$ 7\\
9841.3 $\pm$ 1.3&	1$^+$&	566 $\pm$ 52&	154 $\pm$ 14\\
9871.7 $\pm$ 1.2&	1$^+$&	223 $\pm$ 21&	60 $\pm$ 6\\
9893.9 $\pm$ 1.4&	1$^+$&	116 $\pm$ 12&	31 $\pm$ 3\\ 
10020.5 $\pm$ 1.7&	1$^+$&	71 $\pm$ 12&	18 $\pm$ 3\\ 
10033.9 $\pm$ 1.4&	1$^+$&	210 $\pm$ 25&	54 $\pm$ 6\\\hline
\rule{0pt}{3ex}&&&			$B(E1\uparrow$) (10$^{-3}$  e$^2$ fm$^2$)\\\cline{4-4}
9839.7 $\pm$ 1.2&	1$^-$&	875 $\pm$ 78&	2.6 $\pm$ 0.2\\
9849.3 $\pm$ 1.4&	1$^-$&	6527 $\pm$ 530&	19.6 $\pm$ 1.6\\
9943.3 $\pm$ 1.5&	1$^-$&	123 $\pm$ 18&	0.4 $\pm$ 0.1\\
9951.2 $\pm$ 1.2&	1$^-$&	470 $\pm$ 44&	1.4 $\pm$ 0.1\\
10091.2 $\pm$ 1.4&	1$^-$&	601 $\pm$ 67&	1.7 $\pm$ 0.2\\
10152.3 $\pm$ 1.7&	1$^-$&	1463 $\pm$ 193&	4.0 $\pm$ 0.5\\
10179.7 $\pm$ 1.7&   1$^-$&  1340 $\pm$ 442&3.6 $\pm$ 1.2\\
\end{tabular}
\end{table}   
                                                                            
Table I presents our results for $\Gamma_0$, $B(M1\uparrow)$ and $B(E1\uparrow)$.  
There, level energies $\omega$ rather than $\gamma$-ray transition energies $E_\gamma$ are given. In addition to each level's statistical fit uncertainty, a 1 keV uncertainty was assessed for the HPGe detector energy calibration.
As stated above, prior to the present work only the M1 transition of $E_\gamma$=9757 keV was known \cite{Li}. 
This state and its associated strength were confirmed in the present experiment. 
In addition to this transition seven M1 states were identified, with one at $\omega$=9841.3 keV of similar strength. The combined strength of the remaining six states is comparable to the strength of these two dominant states.
The E1 states and associated strengths are in good agreement with those of Moreh {\it et al.} \cite{Moreh}, except for the 9840 keV state, 
which is identified in the present work as a parity doublet of $B(E1\uparrow)$ at $\omega$=9839.7 keV and $B(M1\uparrow)$ at $\omega$=9841.3 keV, 
with total elastic width $\Gamma_0$ as found in Ref. \cite{Moreh}. 


The main uncertainties associated with the present experimental data are caused by uncertainties in photon flux, $\gamma$-ray detection efficiency, 
and assumptions made in determining the elastic scattering width $\Gamma_0$. 
Statistical uncertainties were smaller for all the transitions observed. 
A 15\% uncertainty was assigned to the total M1 strength of $B(M1\uparrow)=651(98) \times 10^{-3} \mu_N^2$ found in the present experiment in the 
$^{40}$Ar excitation energy range between 9700 keV and 10200 keV. 

\section{Shell-model calculations of the neutral-current neutrino cross section }

At SN energies, NC neutrino-induced nuclear transitions in \textsuperscript{40}Ar are dominated by allowed transitions from the $J^{\pi}=0^+$ ground state to $J^{\pi}=1^+$ excited states \cite{Langanke}. 
This allows us to use the energies of $1^+$ states in \textsuperscript{40}Ar determined in the present work 
and by Gayer \textit{et al.} \cite{Gayer}, together with detailed nuclear structure calculations,  
to calculate a low-energy approximation to the cross section for the inelastic scattering event \textsuperscript{40}Ar($\nu(\bar\nu),\nu'(\bar\nu')$)\textsuperscript{40}Ar\textsuperscript{*}.

The simplest approach is to follow the method used in Ref. \cite{Langanke}, which assumes (1) that all M1 transitions are entirely isovector spin-flip in nature, so that $B(M1\uparrow)$ is directly related to the Gamow-Teller strength ($B(GT)$), 
and (2) that the leading-order, Gamow-Teller operator ($\bf{\sigma}\tau_3$) dominates in the expansion of the nuclear four-current density.
To leading order in initial neutrino energy, $E$, the NC neutrino cross section for a $0^+\rightarrow1^+$ excitation of energy $\omega$ is
\begin{eqnarray}
\sigma=G_F^2V_{ud}^2(E-\omega)^2B(GT)/\pi \, ,
\label{eqL}
\end{eqnarray}
which is then summed over all 1$^+$ excitations to obtain a total cross-section prediction. 
Although this ``M1'' approximation is good at low SN energies, it omits all higher-order operators 
that determine the cross section, and, for example, the interference terms distinguishing neutrinos from antineutrinos are not included.

The formalism for calculating the {\it all-orders}
neutrino- and antineutrino-nucleus interaction cross sections, assuming knowledge of the structure of the states involved, is given by O'Connell, Donnelly and Walecka (ODW) \cite{ODW}.
To implement our second approach, we used the ODW formalism, with
the experimentally determined excitation energies of the 1$^+$ states in $^{40}$Ar and a state-dependent scaling
of the one-body density matrix elements from our shell-model calculations, that  reproduces the observed  $B(M1\uparrow)$.

We carried out six different shell model calculations for M1 distributions in $^{40}$Ar using OXBASH \cite{21} with effective interactions {\it sdpfnow} \cite{num}, {\it sdpfu} \cite{now} and {\it sdpfmu} \cite{uts}. 
We used two model spaces: (1) valence protons are in {\it sd} shells, neutron {\it sd} shells are completely filled and two additional valence neutrons are in {\it f}7/2 and {\it p}3/2; (2) valence protons are in {\it sd} shells and, in addition to filled neutron {\it sd} shells, two valence neutrons are in {\it fp} shells.
In all models, we found that the predicted $B(M1\uparrow)$ for states above 5 MeV
were predominantly spin-flip, but
below 5 MeV, the {\it orbital} $1^+$ operator matrix elements interfered with the Gamow-Teller strengths.

\begin{figure}
\centering
\includegraphics [width = 12cm] {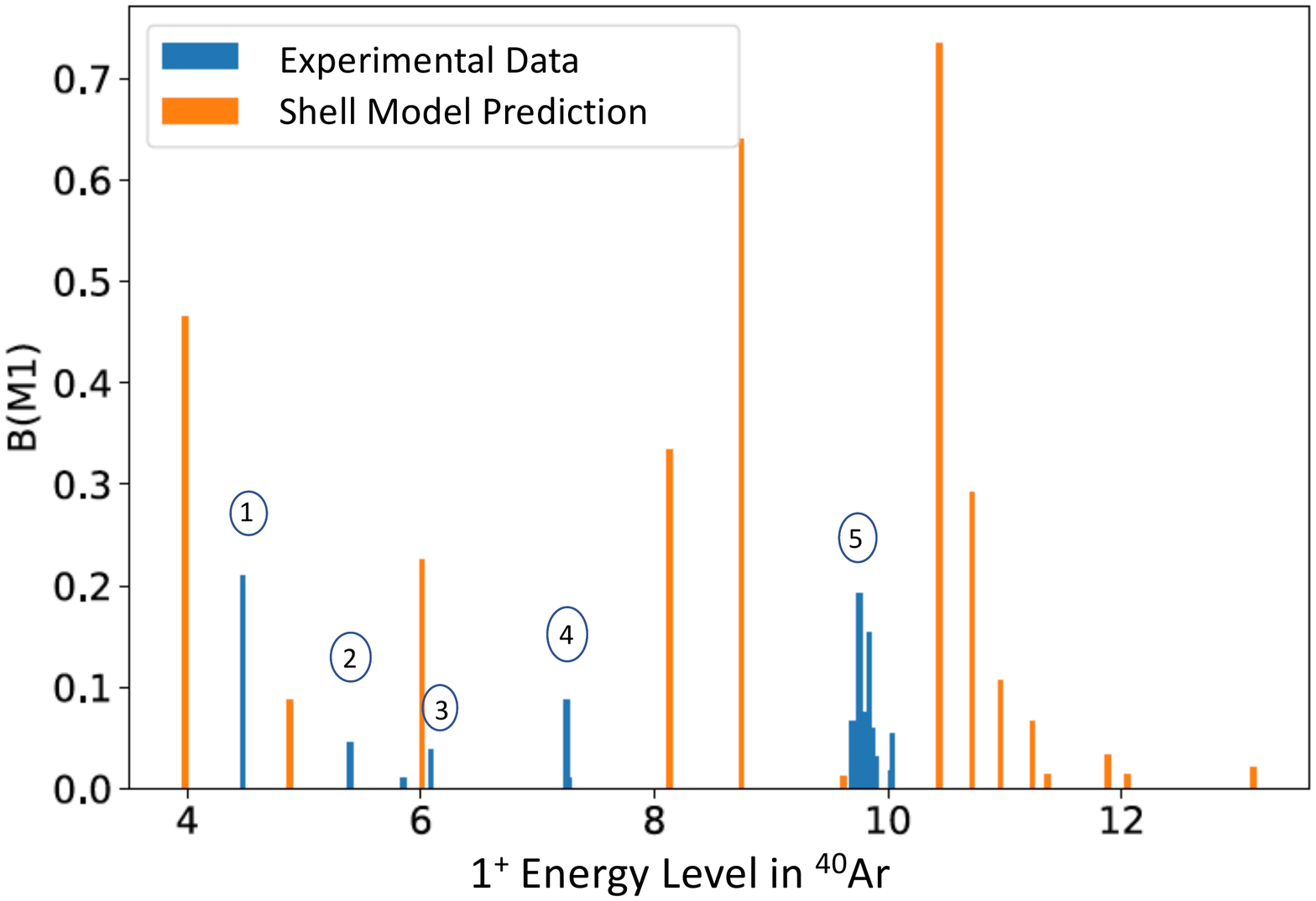}
\caption{The observed and predicted M1 states in $^{40}$Ar obtained using the Utsuno interaction \cite{uts}. 
To obtain a neutral current cross-section prediction from these M1 states, we   
grouped the 14 experimental states into 5 main states, as labelled in the figure.
The shell-model states close in energy to each of the 5 states were also grouped and the
one-body density matrix elements (OBDMEs) scaled to reproduce the 5 (summed) experimental B(M1) values.
Using these scaled OBDMEs, we calculated the cross section using the full  formalism of ODW \cite{ODW}.}
\label{BM1}
\end{figure}

We observed that scaling the nuclear structure one-body density matrix elements for each of the different shell model calculations to reproduce
the experimental energy and B(M1) value 
for every state separately led to a set of cross sections 
that was quite dependent on the model. 
However, we found that if we grouped the 14 1$^+$ states to make up 5 main states, the model dependence of the predicted cross section was reduced considerably.
For this grouping, we chose the 5 states close in energy to the experimental states that exhibit strong M1 strengths, 
with chosen excitation energies 4.473, 5.393, 6.085, 7.25, 9.8 MeV, as summarized in Fig. \ref{BM1}. 
For each of these 5 states, we defined the B(M1) strength to be the sum of the observed experimental 
strengths close to that energy. 
The shell-model states close in energy to each of the 5 states were also grouped and the 
one-body density matrix elements scaled to reproduced the 5 (summed) B(M1) values.

\begin{figure}[b]
\center{  
\includegraphics[width=10 cm]{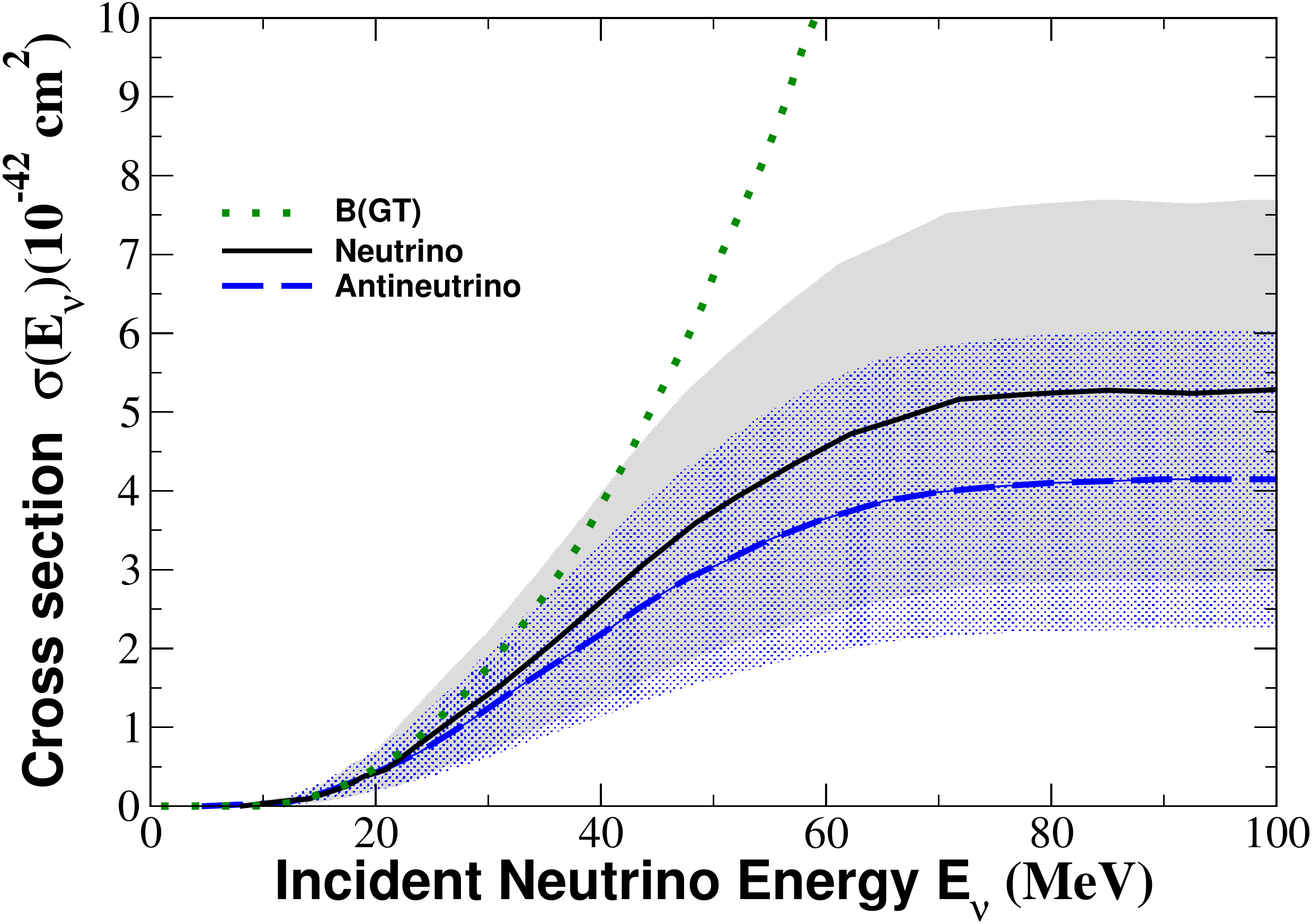}}
\caption{\label{fig:my-label2}The all-orders, neutral current, neutrino (solid) and antineutrino (dashed) cross sections for $0^+\rightarrow1^+$ reactions with \textsuperscript{40}Ar nuclei obtained using the M1 data from the present work and from \cite{Gayer} and compared to the leading-order ``M1'' cross section (dotted). Uncertainties (gray shaded neutrino, blue shaded antineutrino) $\sim 43\%$ were obtained from a combination of the experimental and the shell-model uncertainties.}
\label{secondapprox}
\end{figure}

The resulting predicted cross sections are shown in Fig. \ref{secondapprox} alongside the ``M1'' approximation in Eqn. \ref{eqL}.
As expected, our two approaches give very similar predictions for the NC cross section as $E-\omega$ approaches zero. 
Above about 20 MeV, contributions from operators at higher order in momentum suppress
the predicted neutrino and antineutrino cross sections, while causing the neutrino and antineutrino cross sections to diverge from one another. 
It is straightforward to confirm this deviation from the ``M1" approximation for a general nuclear transition by expanding the necessary operators appearing in the ODW formalism as a function of momentum transfer, using the tables of Donnelly and Haxton \cite{DH}.
Our estimate of the uncertainty on the predicted neutrino and antineutrino cross sections is 43\%, which comes from a combination of the 15\% uncertainty in the observed B(M1) values and an approximate 40\% uncertainty in the shell-model predictions above 5 MeV.
The latter shell-model uncertainty estimate is an estimate based on the differences seen between the scaled ungrouped and the grouped shell-model predictions. 
However, we note that the predictions of the individual shell model spaces and two-body interactions are likely considerably more uncertain.


\section{Conclusion}

In summary, an NRF experiment was performed with a monoenergetic and linearly polarized photon beam of $E_\gamma=9.88$ MeV. 
Photons are uniquely suited to excite dipole states in 0\textsuperscript{+} ground-state nuclei. 
The use of a linearly polarized photon beam allows for an  unambiguous distinguishing between E1 and M1 de-excitation $\gamma$ rays, while the monoenergetic nature of the beam provides a straightforward way to determine potential contributions to the dipole strength from inelastic $\gamma$-ray transitions. Both features are excluded when using Bremsstrahlung beams in NRF experiments. 
The strength and character of 15 dipole states were determined in the 500 keV energy range between 9700 keV and 10200 keV. 
In addition to the already known M1 $\gamma$-ray transition with $E_\gamma=$ 9757 keV \cite{Li}, seven new M1 transitions were identified in the present work. Most notable, the M1 state at 9841.3 keV with $B(M1\uparrow)=(149\pm14)\times10^{-3}\mu_N^2$, which was first reported in \cite{Moreh}, but without any parity assignment. The total M1 strength found in this work is $B(M1\uparrow)=651(98)\times10^{-3}\mu_N^2$. Compared to the previously reported M1 strength of $148(59)\times10^{-3}\mu_N^2$, an increase by almost a factor of 3.4 is observed, resulting in a corresponding increase in the neutral-current supernova neutrino interaction cross section of \textsuperscript{40}Ar in the energy regime accessible with the large LAr detectors at DUNE \cite{DUNE} and the existing detectors IKARUS \cite{IKARUS}, ArgoNeuT \cite{ArgoNeuT}, and the DUNE prototype detectors MicroBOONE \cite{microboone} and ProtoDUNE \cite{protoDUNE}.

The close relationship between the spin-isovector component of the M1 and the Gamow-Teller transition probabilities was used to calculate the neutral current neutrino and antineutrino cross sections for reactions with $^{40}$Ar.
The predicted cross sections are based on shell-model calculations that were scaled to reproduce our measured M1 strengths and 1$^+$ excitation energies. The leading-order approximation is found to be accurate for neutrinos at low supernova energies (up to about 20-30 MeV). The inclusion of operators at all-orders in momentum transfer suppresses both and distinguishes between the neutrino and antineutrino cross sections above 20 MeV.


\section*{Acknowledgements}

H. G. D. Y. is grateful to the Theoretical Division at Los Alamos National Laboratory for funding his visit, during which
he carried out much of the theoretical work presented here.
X. W. wishes to thank the support by the National Natural Science
Foundation of China under Grant Nos. 12275081, U2067205, and U1732138.
The authors are grateful to G. T. Garvey for suggesting this work. They also thank M. Bhike, U. Gayer, B. L{\H o}her and V. Werner for their contributions. The HPGe detector \#3 used in the present work was on loan from the IKP, TU Darmstadt. This work was supported in part by the U.S. Department of Energy (DOE), Office of Nuclear Physics, under Grant No. DE-FG02-97ER41033. The work of A. P. T. is performed in part under the auspices of DOE by LLNL under contract DE-AC52-07NA27344.




\begin{thebibliography}{00}

\bibitem{DUNE}
R. Acciarri {\it et al.},  	arXiv:1601.05471 [physics.ins-det]. 
\bibitem{SURF}
https://sanfordlab.org/
\bibitem{Garcia}M. Bhattacharya, C. D. Goodman, and A. Garcıa, Phys. Rev. C 80 (2009) 055501, and references therein.
\bibitem{Li}
T. C. Li, N. Pietralla, A. P. Tonchev, M. W. Ahmed, T. Ahn, C. Angell, M. A. Blackston, A. Costin, K. J. Keeter, J. Li \textit{et al.}, Phys. Rev. C 73 (2006) 054306.
\bibitem{Gayer}
U. Gayer, T. Beck, J. Isaak, N. Pietralla, P. C. Ries, M. Schilling, V. Werner, D. Savran, M. Bhike, and W. Tornow, Phys. Rev. C 100 (2019) 034305.
\bibitem{Langanke} K. Langanke, G. Martınez-Pinedo, P. von Neumann-Cosel, and A. Richter, Phys. Rev. Lett. 93 (2004) 202501.
\bibitem{NRF2}
U. Kneissl, J. Margraf, H. H. Pitz, P. von Brentano, R. -D. Herzberg and A. Zilges. Prog. Part. Nucl. Phys. 34 (1995) 285.
\bibitem{NRF1}
N. Pietralla, Z. Berant, V. N. Litvinenko, S. Hartman, F. F. Mikhailov, I. V. Pinayev, G. Swift, M. W. Ahmed, J. H. Kelley, S. O. Nelson \textit{et al.}, Phys. Rev. Lett. 88 (2001) 012502.
\bibitem{Rus13} G. Rusev et al., Phys. Rev. Lett. 104 (2013) 072501.
\bibitem{Ton17}  A.P. Tonchev et al., Phys. Lett. B 773 (2017) 20.
\bibitem{HIGS}
H. Weller M. W. Ahmed, H. Gao, W. Tornow, Y. K. Wu, M. Gai, R. Miskimen, Prog. Part. Nucl. Phys. 62 (2009) 257.
\bibitem{TUNL}
http://www.tunl.duke.edu/.
\bibitem{fission}
C. Bhatia, B. Fallin, M. E. Gooden, C. R. Howell, J. H. Kelley, W. Tornow, C. W. Arnold, E. M. Bond, T. A. Bredeweg, M. M. Fowler \textit{et al.}, Nucl. Instr. and Meth. Phys. Res. A 757 (2014) 7.
\bibitem{ring}
P. Ring and P. Schuck, \textit{The Nuclear Many Body Problem}, Springer-Verlag, New York (1980). 
\bibitem{Hamilton}
W. D. Hamilton, editor, \textit{The Electromagnetic interaction in nuclear spectroscopy}, North-Holland Publishing Company (1975).
\bibitem{MCNP}
T. Goorley, "MCNP6.1.1-Beta Release Notes", LA-UR-14-24680 (2014).
\bibitem{Moreh}
R. Moreh, W. C. Sellyey, D. C. Sutton, R. Vodhanel, and J. Bar-Touv, Phys. Rev. C 37 (1988) 2418.
\bibitem{ODW}
J. S. O'Connell, T. W. Donnelly and J. D. Walecka, Phys. Rev. C 6 (1972) 719.
\bibitem{21}
B. A. Brown \textit{et al.}, MSU-NSCL Report No. 1289 (2004).
\bibitem{num}
S. Nummela, P. Baumann, E. Caurier, P. Dessagne, A. Jokinen, A. Knipper, G. Le Scornet, C. Miehé, F. Nowacki, M. Oinonen \textit{et al.}, Phys. Rev. C 63 (2001) 044316.
\bibitem{now}
F. Nowacki and A. Poves, Phys. Rev. C 79 (2009) 014310.
\bibitem{uts}
Y. Utsuno, T. Otsuka, B. A. Brown, M. Honma, T. Mizusaki, N. Shimizu, Phys. Rev. C 86 (2013) 051301(R).
\bibitem{DH}
T. W. Donnelly and W. C. Haxton, At. Data Nucl. Data Tables 25 (1980) 1-28.
\bibitem{IKARUS}
C. Rubbia, M. Antonello, P. Aprili, B. Baibussinov,
M. B. Ceolin, L. Barzè, P. Benetti, E. Calligarich, N.
Canci, F. Carbonara {\it et al.}, 
J. Instrum. 6 (2011) P07011.
\bibitem{ArgoNeuT}
C. Anderson, M. Antonello, B. Baller, T. Bolton, C. Bromberg,
F. Cavanna, E. Church, D. Edmunds, A. Ereditato, S. Farooq
{\it et al.}, J. Instrum. 7 (2012) P10019.
\bibitem{microboone}
R. Acciarri {\it et al.} (The MicroBooNE Collaboration), J. Instrum. 12 (2017) P02017.
\bibitem{protoDUNE}
The DUNE Collaboration {\it et al.}, J. Inst. 17 (2022) P01005.

\end{thebibliography}


\end{document}